\documentclass[preprint,12pt]{elsarticle}
\biboptions{square,numbers,sort&compress}
\usepackage{amsmath}
\usepackage{algorithmic}
\usepackage{array}
\usepackage{mdwmath}
\usepackage{mdwtab}
\usepackage{eqparbox}
\usepackage{fixltx2e}
\usepackage{stfloats}
\usepackage{amssymb}
\usepackage{amsthm}

\usepackage{float}
\usepackage{fixltx2e} 
\usepackage{epsfig,color}
\usepackage{subfig}
\usepackage{yhmath}
\usepackage{stackrel}

\journal{arXiv.org}
\begin{document}


\begin{frontmatter}


\title{On the Definition and Existence of an MVU Estimator for Target Location Estimation \tnoteref{t1}}
\tnotetext[t1]{We would like to honor and acknowledge the memory of Professor Alireza Seyedi who had an inspiring role for this work and passed away in Oct 2014 before the preparation of this manuscript. I would also like to thank Professor Carl Mueller, Professor Michele Gage and Professor Mark Bocko for helpful discussions and invaluable feedbacks. In addition, my appreciation goes out to Mona Komeijani for preparing the illustrations of this paper. }

\author[label1]{Arian~Shoari\fnref{fn1}}
\ead{shoari@ece.rochester.edu}
\fntext[fn1]{Corresponding author}

\address[label1]{Department of Electrical and Computer Engineering, University of Rochester, Rochester, NY, USA}

\begin{abstract}
The problem of target localization with ideal binary detectors is considered in one dimensional space. The problem is investigated in both a censored and non-censored scheme. In the censored setting, the problem is equivalent to estimating the center of a uniform distribution by knowing samples of data. It does not admit an MVU estimator according to the previous results of Lehmann-Sheffe. However, it is proven that if the radius of detection is known and sensor deployment region is very large, both censored and non-censored cases will have an MVU estimator among the functions that are invariant to Euclidean motion. In addition, it is shown that when the radius of detection is not known, the censored case still has an MVU estimator whereas in the non-censored case, an MVU estimator does not exist,  even under the assumption that the estimators are invariant to Euclidean motion.
\end{abstract}

\begin{keyword}
MVU Estimator \sep Completeness  \sep Minimal Sufficient Statistics \sep Lehmann-Sheffe theorem \sep Bondesson theorem \sep Localization
\end{keyword}

\end{frontmatter}

\section{Introduction}
\label{sec:Introduction}
The problem of localization of a target using a number of distributed measurements has been addressed widely in the electrical engineering literature \citep{Amundson09,Wang2012}. The problem originally was considered in radar and sonar \citep{Weinstein82Sonar,Carter81Sonar} but there is renewed interest in it due to advances in wireless sensor networks \citep{Chen2002ApplicationsofWSN,Chen2003ApplicationsofWSN,Karl2005WSN} and applications in 4G/ LTE networks \citep{stefania2009lte,Mohapatra2005LocBasedServ,Yang2012SecuringLoc}. The localization usually is performed through measurements such as Time Difference of Arrival (TDOA)\citep{Kung1998TDOA,Yang2005TDOA,Yang2006TDOA2}, Direction of Arrival (DOA)\citep{Oshman1999DOA,Kaplan2001DOA,Peng06AoA}, Frequency Difference of Arrival (FDOA)\citep{Lee2007,Yu2012,Fuyong2014comments}, and Radio Signal Strength (RSS)\citep{sheng2003collaborative,Sheng05,Blatt06,kieffer2006centralized,Vaghefi11_RSSbased,Rabbat2005,Ampeliotis08,Bulusu02,VarshneyFading07,VarshneyBinCoh,Arian2014SSP,ArianSpawc2010,Liu2007Target,Murthy2011multiple,Murthy2012multiple}. The first three methods do not fit well within energy and complexity constraints of wireless sensor networks especially when the target is not cooperative \citep{sheng2003collaborative,Liu2007Target}. Examples of applications when the target does not assist in evaluating its position are identifying the primary user in cognitive radio\citep{ArianSpawc2010}, spectrum cartography \citep{Mateos2009}, identify unauthorized users of bandwidth and localizing jammers positions in the battlefield \citep{Arian2012Milcom}. Although measurements of received signal strength can be exact, it is not practical to transmit a non-quantized measurement result to the fusion center (FC), where the decision about the location of a target is reached because unlimited bandwidth is needed. Thus, in many papers it is assumed that the sensors make a quantized \citep{VarshneyFading07} or binary measurement \citep{zhou2009smallest,VarshneyBinCoh,Arian2014SSP,ArianSpawc2010,Liu2007Target,Murthy2011multiple,Shrivastava_2009_Target_Tracking,Murthy2012multiple} of the received signal strength. Binary measurements are preferred because the implementation is simple and requires minimum bandwidth. In addition, in localization using binary sensors, some papers assume that only the detecting sensors (which generate a "one" output) will communicate data to FC \citep{zhou2009smallest,ArianSpawc2010,Antonio04,Murthy2012multiple}. This censoring scheme saves communication overhead and energy because if the sensor deployment region is very large it is expected that most sensors do not detect the target \citep{artes2004target}.

From a mathematical point of view, the problem can be considered as a point-wise estimation when samples of data are available from a population function representing the probability of detection at a specific distance from the target. The case when measurements are noise-free is a limiting scenario and its performance can be considered as the lower bound for all other localizers with uncertain measurements such as those in the presence of noise and fading. Thus, analysis of this scenario is important. In this case, the problem is equivalent to estimating the centroid of a uniform distribution from a limited number of samples. However, it appears that this formulation of the problem does not have a Cramer Rao Bound (CRB) because the likelihood function is not well behaved \citep{Kay93Estimationbook}. Therefore, a reasonable approach to establish the lower bound for the performance in this scenario would be to find an MVU estimator. This raises the question if a minimum variance unbiased estimator exists at all. One standard approach to this problem is to first examine the existence of complete sufficient statistics \citep{Kay93Estimationbook}. The relation between completeness and existence of a uniform minimum variance unbiased estimator has long been of interest to Statisticians. For example  Lehmann-Sheffe show that if a complete sufficient statistics exists, all estimable parametric functions could be uniformly MVU estimated \citep{lehmann1950completeness}. Unfortunately, in this case a complete sufficient statistics does not exist most of the time as we will see in Sections \ref{sec:CensoredRKnown} and \ref{sec:OntheExistanceNonCensored}. Bahadur in \citep{bahadur1957unbiased} provides a converse to the Lehmann-Sheffe theorem, and show that if every estimable parametric function admits a UMVUE (uniform minimum variance unbiased estimator), then a complete sufficient statistic exists. However, \citep{bondesson1983uniformly,rao1973linear} provide examples to show that even when a complete sufficient statistics does not exist, there may still be a chance that an MVU estimator exists for some parametric function. In Section \ref{sec:CensoredRKnown}, we will mention some known mathematical results that suggests that an MVU estimator does not exist for this problem in the censored setting. However, we provide a constraint on the function space which seems necessary in location estimation and prove that under that constraint, the MVU estimator exists. We believe that this constraint is critical for many applied estimators and should be considered in the  evaluation of MVU estimators in many other applications. We also consider  different cases of this  problem and discuss the existence of MVU estimators for each of them. We believe that this work also may have pedagogical value in the context of statistical inference.

In Section \ref{sec:ProbFormulation} we will formalize the problem and mathematical notions. In Section \ref{sec:OntheExistanceCensored} we will consider the problem in a censored scenario. In Section \ref{sec:OntheExistanceNonCensored} we will discuss the results for a non-censored case. Concluding remarks and discussion is presented in Section \ref{sec:Conclusion}.

\section{Problem Formulation}
\label{sec:ProbFormulation}
Assume that a target is located at an unknown location $z_{\text{T}}$ in space and transmits a signal whose power propagates isotropically and is attenuated monotonically as a function of distance from the target. To simplify matters, we solve the problem in one dimensional space, an approach that has been widely studied in the statistics literature. $N$ sensors, are randomly scattered in a deployment region, $\mathbb{G}=[A_1,A_2]$ of length $L=A_2-A_1$. They measure the received power and compare it to a threshold, $\tau$, to make a binary decision about the target presence.
Let $n$ be the number of detecting sensors (the ones that report decision "one" to FC) and $p$ be the number of non-detecting sensors (the ones whose decisions are "zero"). Also let ${z_\text{D}}_i$ and ${z_\text{ND}}_j$ represent the locations of the $i$th detecting and $j$th non-detecting sensors respectively. We represent the set of indices of all detecting sensors by $\mathcal{S}_{\text{D}}=\{1,..,n\}$ and the set of indices of all non-detecting sensors by $\mathcal{S}_{\text{ND}}=\{1,..,p\}$. Therefore, $\mathbf{Z}_\text{D}=\left[{z_\text{D}}_i | i \in \mathcal{S}_{\text{D}} \right]$ is a vector containing the locations of all detecting sensors and
$\mathbf{Z}_\text{ND}=\left[{z_\text{ND}}_j  | j \in \mathcal{S}_{\text{ND}} \right]$ is a similar vector for non-detecting sensors.

We assume the sensors make a noise free decision, which can be considered as the  limiting case when the measured power is averaged over a sufficiently long duration of time. Since the received power is a decreasing function of distance from the target, this means that all the detecting sensors are located within a detection radius, $R$, from the target i.e. $\forall i \in \mathcal{S}_{\text{D}}, {z_\text{D}}_i \in [z_{\text{T}}-R,z_{\text{T}}+R]$. $R$ can be considered known or unknown depending on whether or not the propagation model and the transmit power are known to FC; we will investigate each case separately.
We also assume that at least one sensor is a detecting sensor ($n\ge1$) and $\mathbb{G}$ is sufficiently large such that $[z_{\text{T}}-3R,z_{\text{T}}+3R] \subset [A_1,A_2]$. Hence, the results of this work are also valid when $\mathbb{G}$ is unlimited (entire $x$ axis). In addition we assume that at least one sensor detects the target.

We will consider the problem both in a censored and non-censored scheme. In a censored scenario presented in Section \ref{sec:OntheExistanceCensored}, we assume the sensors are configured such that only the detecting sensors report their locations to the FC. Whereas, in the non-censored scenario discussed in Section \ref{sec:OntheExistanceNonCensored} both detecting and non-detecting sensors inform FC of their locations and decisions.

\newpage
\section{On the Existence of MVU Estimator in the Censored Scheme}
\label{sec:OntheExistanceCensored}
In this scenario, only the locations of detecting sensors are available to FC. Therefore, the localization problem is considered as a special instance of point estimation and can be viewed as estimating the center of a uniform distribution by knowing samples of data from the uniform distribution.

We divide the problem into the cases in which the radius of detection, $R$, is known to FC and when it is unknown and we investigate each one separately.

\subsection{Known Detection Radius}
\label{sec:CensoredRKnown}

In this case, $R$ is known and the goal is to estimate $ z_{\text{T}}$ from $n$ samples of a uniform distribution over $\left[ z_{\text{T}}-R,z_{\text{T}}+R\right]$. It is known in the statistics literature that  $({z_\text{D}}_{L},{z_\text{D}}_{U})$ in which ${z_\text{D}}_{L}=\min_{i \in \mathcal{S}_{\text{D}}}  {z_\text{D}}_i $  and ${z_\text{D}}_{U}=\max_{i \in \mathcal{S}_{\text{D}}}  {z_\text{D}}_i$ is a minimal sufficient statistics for estimation of $z_{\text{T}}$ \citep{lehmann1981interpretation,David2003OrderStatistics}. The intuition behind is that if the minimum and maximum of the samples are known, the rest of the data will not provide any information about the target because any sensor in between must be a detecting sensor. For the sake of notational simplicity, let us denote ${z_\text{D}}_{L}$ and ${z_\text{D}}_{U}$ by $x$ and $y$ respectively in this sub-section. If we consider any estimator of $z_{\text{T}}$, as $\hat{ z_{\text{T}}}=v(\mathbf{Z}_\text{D})$, the Rao-Blackwell theorem indicates that $h(x,y)=E[v(\mathbf{Z}_\text{D})\mid (x,y)]$ would have a smaller or equal variance to that of $v(\mathbf{Z}_\text{D})$ where $E$ represents the expectation of the random variable \citep{Kay93Estimationbook}. Thus, we can restrict our search for an MVU estimator to the functions of $(x,y)$.

\begin{figure}[!t]
\begin{centering}
\includegraphics[width=5 in]{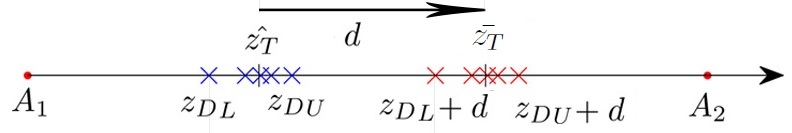}
\caption{Non-invariance to Euclidian motion: $\bar{z_{\text{T}}}=\hat{z_{\text{T}}}+d$}
\label{Non-InvarianceEuclidian}
\end{centering}
\end{figure}

Although $(x,y)$ are a minimal sufficient statistics for this problem, a complete sufficient statistics does not exist \citep{lehmann1981interpretation,David2003OrderStatistics}. For the case of a uniform distribution,  \citep{lehmann1950completeness} proved that no non-constant parametric function of $z_{\text{T}}$ (including $z_{\text{T}}$) can admit a Uniform Minimum Variance Unbiased estimator (UMVUE). \citep{bondesson1975uniformly} provided another proof for the nonexistence of UMVUE for this problem which is true even if $n=1$ (estimators built based on one observation). However, in location estimation the estimators have characteristics that limit the function space and a Minimum Variance Unbiased estimator may exist among those functions. For example, in location estimation (and perhaps many other estimators of physical phenomenon) it is reasonable to assume that estimators are invariant under Euclidean motion i.e. we expect that if the  observations shift, then the estimate of the phenomenon shifts along with the observations as illustrated in Figure \ref{Non-InvarianceEuclidian} . Therefore, if we estimate $z_{\text{T}}$ by $h(x,~ y)$, we should have
\begin{eqnarray}
h(x+d,y+d)=h(x,y)+d.
\end{eqnarray}

\begin{figure}[!t]
\begin{centering}
\includegraphics[width=5 in]{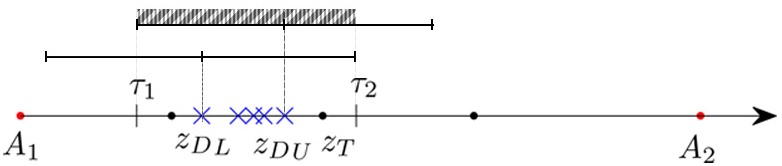}
\caption{\textit{Possible target region} ilustration for censored scenario, $\tau_1={z_\text{D}}_{U}-R$ and $\tau_2={z_\text{D}}_{L}+R$. }
\label{ProblemSettingsandT}
\end{centering}
\end{figure}

Let us define the \textit{possible target region} for $\mathbf{Z}_\text{D}$ as $\mathcal{T}(\mathbf{Z}_\text{D})=\left[y-R,x+R\right]$ as  illustrated by hashed interval in Figure\ref{ProblemSettingsandT}. From the definition, it is clear that $\mathcal{T}(\mathbf{Z}_\text{D})=\mathcal{T}(x,y)$. An observation  $(x,y)$ implies that $z_{\text{T}} \in \mathcal{T}(x,y)$. On the other hand, any observation in which $z_{\text{T}} \in \left[y-R,x+R\right] $ is a possible observation because this along with $x \leq y$ guarantee that  $\left|x -z_{\text{T}} \right| \leq R$ and $\left|y -z_{\text{T}} \right| \leq R$. It can be shown that the $\mathcal{T}(x,y)$ interval and $(x,y)$ are related by a one to one mapping. Thus, in the search for an MVU function it is possible to search on functions of $\mathcal{T}(x,y)$ instead. In other words, the estimator can be limited to,
\begin{eqnarray}
g(\mathcal{T}(x,y)).
\end{eqnarray}
Therefore, the invariance under Euclidean motion suggests that
\begin{eqnarray}
g(\mathcal{T}(x+d,y+d))=g(\mathcal{T}(x,y))+d.
\end{eqnarray}
Now consider a sub-partition of random space that has $n$ detecting sensor. The mean squared error for this sub-partition becomes,
\begin{eqnarray}
E[\left|\hat{z_{\text{T}}}-z_{\text{T}}\right|^2]=  2 \binom{n}{2}  \int_{z_{\text{T}}-R}^{z_{\text{T}}+R}  \int_{z_{\text{T}}-R}^{z_{\text{T}}+R} \left(\frac{y-x}{L}\right)^{n-2}  \left(\frac{1}{L}\right)^2 \left| g(\mathcal{T}(x,y))-z_{\text{T}} \right|^2 \nonumber\\
\mathbf{1}_{x \leq y}   dy dx.
\end{eqnarray}
where $\mathbf{1}_{X}$ is the indicator function of $X$, the binomial coefficient compensates for the fact that any two of the $n$ sensors can be selected as the maximum and minimum values of the data; and $\left(\frac{y-x}{L}\right)^{n-2}$ is the probability that all other sensors are located inside $[x,y]$.
Now, with a change of variable $t=y-x$, removing $y$ and considering that $t \in \left[0,2R \right]$, we have,
\begin{eqnarray}
E[\left|\hat{z_{\text{T}}}-z_{\text{T}}\right|^2]=  n(n-1)  \int_{z_{\text{T}}-R}^{z_{\text{T}}+R}  \int_{0}^{2R} \left(\frac{t}{L}\right)^{n-2}  \left(\frac{1}{L}\right)^2 \left| g(\mathcal{T}(x,x+t))-z_{\text{T}} \right|^2 \nonumber\\
\mathbf{1}_{z_{\text{T}}\in \left[ t+x-R,R+x \right] }  dt dx.
\end{eqnarray}
The assumption regarding invariance of $g$ to Euclidean motion results in $g(\mathcal{T}(0+x,t+x))=g(\mathcal{T}(0,t))+x$. In addition, $\mathbf{1}_{z_{\text{T}}\in \left[ t+x-R,R+x \right]}=\mathbf{1}_{z_{\text{T}}-x \in \left[ t-R,R\right]}$. Thus, we have,
\begin{eqnarray}
E[\left|\hat{z_{\text{T}}}-z_{\text{T}}\right|^2]=  n(n-1)  \int_{z_{\text{T}}-R}^{z_{\text{T}}+R}  \int_{0}^{2R} \left(\frac{t}{L}\right)^{n-2}  \left(\frac{1}{L}\right)^2 \left| g(\mathcal{T}(0,t))+x-z_{\text{T}} \right|^2  \nonumber\\
\mathbf{1}_{z_{\text{T}}-x \in \left[ t-R,R\right] }  dt dx.
\end{eqnarray}
With another change of variable $t'=z_{\text{T}} -x$ and eliminating $x$ and $z_{\text{T}}$ we have,
\begin{eqnarray}
E[\left|\hat{z_{\text{T}}}-z_{\text{T}}\right|^2]=  n(n-1)  \int_{-R}^{+R}  \int_{0}^{2R} \left(\frac{t}{L}\right)^{n-2}  \left(\frac{1}{L}\right)^2 \left| g(\mathcal{T}(0,t))-t' \right|^2   \mathbf{1}_{t'\in \left[ t-R,R\right] }  \nonumber\\
dt dt'.
\end{eqnarray}
Moreover, because the interval of the two integrals are independent, we can change the order of the integrals and obtain,
\begin{eqnarray}
E[\left|\hat{z_{\text{T}}}-z_{\text{T}}\right|^2]= n(n-1)   \int_{0}^{2R}   \left(\frac{t}{L}\right)^{n-2}  \left(\frac{1}{L}\right)^2 \int_{-R}^{+R}  \left| g(\mathcal{T}(0,t))-t' \right|^2   \mathbf{1}_{t'\in \left[ t-R,R\right] }\nonumber\\
dt' dt.
\end{eqnarray}
Furthermore, because $t \ge 0$ we know that $\left[ t-R,~ R \right] \subset \left[  -R, ~ R \right]$, thus,
\begin{eqnarray}
E[\left|\hat{z_{\text{T}}}-z_{\text{T}}\right|^2]&=&  n(n-1)   \int_{0}^{2R}   \left(\frac{t}{L}\right)^{n-2}  \left(\frac{1}{L}\right)^2 \left[  \int_{t-R}^{R}  \left| g(\mathcal{T}(0,t))-t' \right|^2  dt' \right]  dt. \nonumber\\
\end{eqnarray}
Substituting $\mathcal{T}(0,t)$ by its value $[t-R,R]$ we have,
\begin{eqnarray}
\label{LastMSECensored}
E[\left|\hat{z_{\text{T}}}-z_{\text{T}}\right|^2]&=&  n(n-1)   \int_{0}^{2R}   \left(\frac{t}{L}\right)^{n-2}  \left(\frac{1}{L}\right)^2 \left[  \int_{t-R}^{R}  \left| g([t-R,~R])-t' \right|^2  dt' \right]  dt. \nonumber\\
\end{eqnarray}
On the other hand, $g([t-R,~R])$ does not depend on $t'$ and we know that if it is selected as the center of gravity of the inner integration interval i.e.
\begin{eqnarray}
\label{CGCensored}
g([t-R,~R])=CG([t-R,R])=\frac{t-R+R}{2}.
\end{eqnarray}
the inner integral will always be minimized \citep{longley2005geographic,Tan2006DataMining,kogan2006grouping}. Hence because the outer integral integrand is always positive, the entire expectation will be minimized.

Furthermore, similar to derivation presented in (\ref{LastMSECensored}), we can show that
\begin{eqnarray}
\label{LastBiaseCensored}
E[\hat{z_{\text{T}}}-z_{\text{T}}]&=&  n(n-1)   \int_{0}^{2R}   \left(\frac{t}{L}\right)^{n-2}  \left(\frac{1}{L}\right)^2 \left[  \int_{t-R}^{R}  \left( g([t-R,~R])-t' \right)  dt' \right]  dt. \nonumber\\
\end{eqnarray}
Thus, it can be verified that $g([t-R,~R])=CG([t-R,R])=\frac{t-R+R}{2}$ makes (\ref{LastBiaseCensored}) zero i.e. (\ref{CGCensored}) is unbiased. Considering that the following are equivalent
\begin{eqnarray}
g([t+x-R,~x+R])=g([t-R,~R])+x= CG([x+t-R,x+R])
\end{eqnarray}
Then, $g(\mathcal{T}(x,y))=CG(\mathcal{T}(x,y))$ will be the minimum variance unbiased estimator among all estimators that are invariant to Euclidean motion.

\subsection{Unknown Detection Radius}
\label{sec:CensoredRUnknown}
In this sub-section, we will consider the problem when $R$ is unknown which means that  $[z_{\text{T}},R]$ are both parameters of estimation now. In this case, $({z_\text{D}}_{L},{z_\text{D}}_{U})$ not only is minimal sufficient statistics but also is complete \citep{lehmann1981interpretation,David2003OrderStatistics}. Now consider a mean estimator $v(\mathbf{Z}_\text{D})=\frac{1}{n} \sum_{i \in \mathcal{S}_{\text{D}}} {z_\text{D}}_i$. According to to Rao-Blackwell theorem \citep{Kay93Estimationbook}, $\hat{z_{\text{T}}}=E\left[v(\mathbf{Z}_\text{D}) | ({z_\text{D}}_{L},{z_\text{D}}_{U})\right]= \frac{{z_\text{D}}_{L}+{z_\text{D}}_{U}}{2}$ would be the MVU  estimator which is the same result as has been derived in Section \ref{sec:CensoredRKnown} but without any constraint on function space of the estimator and for a different reason.

\section{On the Existence of an MVU Estimator in the Non-Censored Scheme}
\label{sec:OntheExistanceNonCensored}
In this scenario both detecting and non-detecting sensors report their locations and observations to FC.
Therefore, the problem can be viewed as estimation of the center of a uniform distribution over $[z_{\text{T}}-R,z_{\text{T}}+R]$ given its samples and samples of another uniform distribution over the complemented interval $[A_1,A_2]-[z_{\text{T}}-R,z_{\text{T}}+R]$.
We define the following,
\begin{eqnarray}
{z_\text{D}}_{L}&=&\min_{i \in \mathcal{S}_{\text{D}}}  {z_\text{D}}_i \nonumber\\
{z_\text{D}}_{U}&=&\max_{i \in \mathcal{S}_{\text{D}}}  {z_\text{D}}_i \nonumber\\
{z_\text{ND}}_{L}&=&\max_{j \in \mathcal{S}_{\text{ND}} ~\&~ {z_\text{ND}}_j < {z_\text{D}}_{L}}  {z_\text{ND}}_{j} \nonumber\\
{z_\text{ND}}_{U}&=&\min_{j \in \mathcal{S}_{\text{ND}} ~\&~ {z_\text{ND}}_j > {z_\text{D}}_{U}}  {z_\text{ND}}_{j}.
\end{eqnarray}
With the assumption that at least one sensor is a detecting sensor, it's straight forward to show that  ${z_\text{ND}}_{L}$ summarizes the information of all non-detecting sensors located at $z < {z_\text{D}}_{L}$; and  ${z_\text{ND}}_{U}$ summarizes the information of all non-detecting sensors located at $z >{z_\text{D}}_{U}$. In addition, the combination of ${z_\text{D}}_{L}$ and ${z_\text{D}}_{U}$ summarizes the information of all detecting sensors.

\subsection{Known Detection Radius}
\label{sec:NonCensoredRknown}
\begin{figure*}[!t]
\begin{tabular}{c}
\includegraphics[width=5 in]{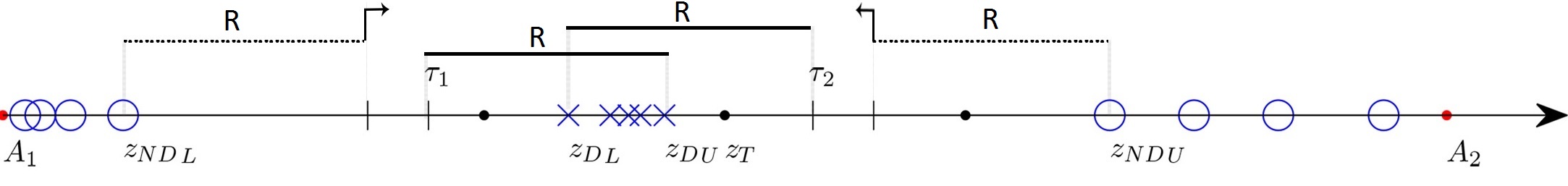}\\
(a)\\
\includegraphics[width=5 in]{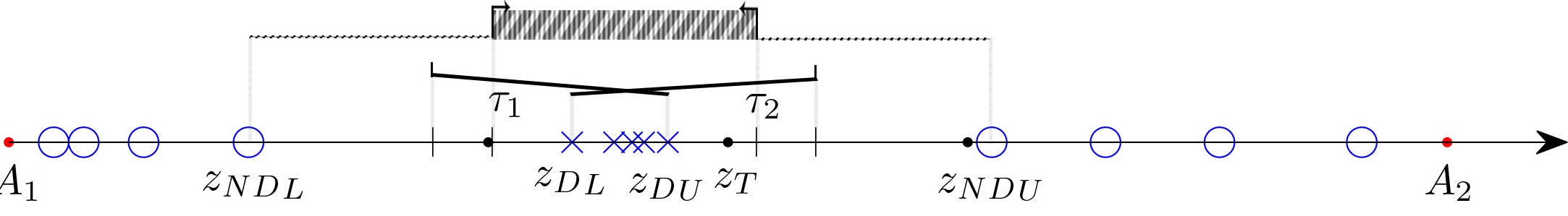} \\
(b)\\
\includegraphics[width=5 in]{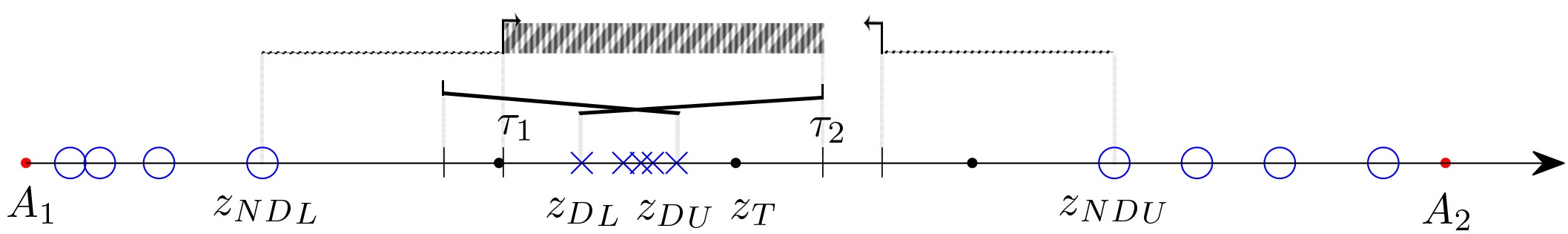} \\
(c) \\
\includegraphics[width=5 in]{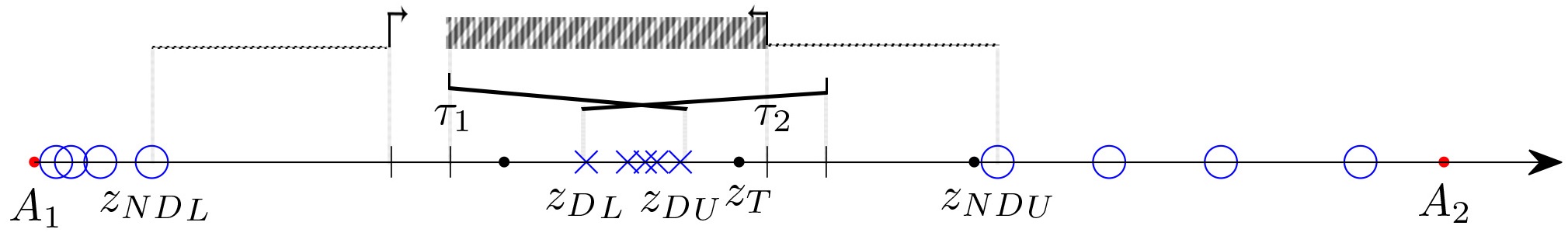} \\
(d)\\
\end{tabular}
\caption{The four possible relative locations of  ${z_\text{D}}_{L}$,${z_\text{D}}_{U}$,${z_\text{ND}}_{L}$,${z_\text{ND}}_{U}$ in non-censored scenario and the effect in $\tau_1$ and $\tau_2$ determination.}
\label{NonDetectingAllConditions}
\end{figure*}

In this scenario, the \textit{possible target region} can be defined as
\begin{eqnarray}
\label{PTRDefDPlusND}
\mathcal{T}(\mathbf{Z}_\text{D},\mathbf{Z}_\text{ND})=\left[{z_\text{D}}_{U}-R,{z_\text{D}}_{L}+R\right]-\left([{z_\text{ND}}_{L}-R,{z_\text{ND}}_{L}+R] \bigcup  \right. \nonumber\\
\left.[{z_\text{ND}}_{U}-R,{z_\text{ND}}_{U}+R]\right)
\end{eqnarray}
The following conditions categorize the possible situation of ${z_\text{D}}_{L}$, ${z_\text{D}}_{U}$, ${z_\text{ND}}_{L}$ and ${z_\text{ND}}_{U}$ with respect to each other as illustrated in Figures \ref{NonDetectingAllConditions}(a)-(d).

\begin{eqnarray}
\label{conditions}
\left\{ \begin{array}{ll}
i)  &  \left({z_\text{D}}_{U}-R\right) \ge \left({z_\text{ND}}_{L}+R\right) \&    \left({z_\text{D}}_{L}+R\right) \le \left({z_\text{ND}}_{U}-R\right)  \\
ii)  &   \left({z_\text{D}}_{U}-R\right) < \left({z_\text{ND}}_{L}+R\right)  \&    \left({z_\text{D}}_{L}+R\right) >  \left({z_\text{ND}}_{U}-R\right) \\
iii)  & \left({z_\text{D}}_{U}-R\right) < \left({z_\text{ND}}_{L}+R\right)  \&    \left({z_\text{D}}_{L}+R\right) \le \left({z_\text{ND}}_{U}-R\right)    \\
iv)  & \left({z_\text{D}}_{U}-R\right) \ge \left({z_\text{ND}}_{L}+R\right)   \&    \left({z_\text{D}}_{L}+R\right) >  \left({z_\text{ND}}_{U}-R\right)  \\
\end{array}\right.
\end{eqnarray}
Depending on which of these conditions are true, $\mathcal{T}(\mathbf{Z}_\text{D},\mathbf{Z}_\text{ND})$ could be calculated based on only two of the elements of $\{ {z_\text{D}}_{L},{z_\text{D}}_{U},{z_\text{ND}}_{L},{z_\text{ND}}_{U}\}$ as follows

\begin{eqnarray}
\label{conditionalT}
\mathcal{T}(\mathbf{Z}_\text{D},\mathbf{Z}_\text{ND})=\left\{ \begin{array}{lcl}
\left[{z_\text{D}}_{U}-R,{z_\text{D}}_{L}+R\right]  & \mbox{if}  &  condition (i)  \\
\left[{z_\text{ND}}_{L}+R,{z_\text{ND}}_{U}-R\right]   & \mbox{if}  & condition (ii) \\
\left[{z_\text{ND}}_{L}+R,{z_\text{D}}_{L}+R\right]   & \mbox{if}  & condition (iii) \\
\left[{z_\text{D}}_{U}-R,{z_\text{ND}}_{U}-R\right]  & \mbox{if}  &  condition (iv) \\
\end{array}\right.
\end{eqnarray}

Let us represent the beginning and end of the interval $\mathcal{T}(\mathbf{Z}_\text{D},\mathbf{Z}_\text{ND})$ by $\tau_1$ and $\tau_2$.
$\mathcal{T}(\mathbf{Z}_\text{D},\mathbf{Z}_\text{ND})=[\tau_1,\tau_2]$ incorporates all information regarding the target location. Moreover, an observation is valid if and only if $z_{\text{T}} \in \mathcal{T}(\mathbf{Z}_\text{D},\mathbf{Z}_\text{ND})$.
It can be shown that, $\mathcal{T}(\mathbf{Z}_\text{D},\mathbf{Z}_\text{ND})$, is a minimal sufficient statistics for estimation of $z_{\text{T}}$ although it is not complete. \footnote{To show the incompleteness we can equivalently consider the minimal sufficient statistics as $(\tau_1,(\tau_2-\tau_1))$. In this way, $(\tau_2-\tau_1)$ is an ancillary statistics whose PDF does not depend on $z_{\text{T}}$ but in combination with $\tau_1$ provides information about $z_{\text{T}}$. Therefore, the minimal sufficient statistics fail to rid itself from ancillary part of data which might be considered as an indication that it is not complete \citep{lehmann1981interpretation}.}

Moreover, the probability density function that certain detecting sensors are located at $\mathbf{Z}_\text{D}$ and non-detecting sensors are located at $\mathbf{Z}_\text{ND}$, can be defined as
\begin{eqnarray}
f(\mathbf{Z}_\text{D},\mathbf{Z}_\text{ND})=\left(\frac{1}{L}\right)^{n+p} \mathbf{1}_{z_{\text{T}} \in \mathcal{T}(\mathbf{Z}_\text{D},\mathbf{Z}_\text{ND})}
\end{eqnarray}
Because the four conditions in (\ref{conditionalT}) can not be simultaneously true for any observation, we can partition the observation space based upon which condition is met. We will proceed and show that $g=CG\left(\mathcal{T}(\mathbf{Z}_\text{D},\mathbf{Z}_\text{ND})\right)$ is unbiased and minimizes the MSE on each of these partitions under the condition that the deployment region is sufficiently large and the estimators are invariant to Euclidean motion. Therefore, $g=CG\left(\mathcal{T}(\mathbf{Z}_\text{D},\mathbf{Z}_\text{ND})\right)$ is an MVU estimator over the entire random space.

We will provide the proof for one of the partitions below. The details of the other three are similar and have been provided in appendix.
We will partition each sub-space of random space further by considering the number of detecting sensors and non-detecting sensors.

\subsubsection{proof for condition iii}

This condition means that $\mathcal{T}(\mathbf{Z}_\text{D},\mathbf{Z}_\text{ND})=\left[{z_\text{ND}}_{L}+R,{z_\text{D}}_{L}+R\right]$ . In this sub-section, let us denote ${z_\text{ND}}_{L}$ and ${z_\text{D}}_{L}$ by $x$ and $y$ respectively. Hence, the \textit{possible target region} would be $\mathcal{T}(\mathbf{Z}_\text{D},\mathbf{Z}_\text{ND})=\mathcal{T}(x,y)=\left[x+R,y+R\right]$. Moreover, we have that
\begin{eqnarray}
{z_\text{D}}_{U} < {z_\text{ND}}_{L}+2R  \nonumber\\
{z_\text{ND}}_{U} \ge {z_\text{D}}_{L}+2R  \nonumber\\
{z_\text{D}}_{U} \ge {z_\text{D}}_{L}
\end{eqnarray}
Therefore, all detecting sensors should be located in interval $[y,x+2R]$ and all non-detecting sensors should be located in interval $[A_1,x]$ or $[y+2R,A_2]$. An $(x,y)$ observation in this case implies that $z_{\text{T}} \in \mathcal{T}(x,y)$. Note that in this case $y<{z_\text{D}}_{U}<x+2R$ along with $z_{\text{T}} \in \left[x+R,y+R\right]$ guarantees that $z_{\text{T}} -3R<x< z_{\text{T}} -R$ and $(z_{\text{T}}-R)<y<(z_{\text{T}}+R)$.

As explained in Section \ref{sec:CensoredRKnown}, we can restrict our search for an MVU estimator to functions of the minimal sufficient statistics i.e. $g(\mathcal{T}(\left(x,y\right)))=g(\left[x+R,y+R\right])$. The invariance to Euclidean motions result in $g(\mathcal{T}(x+d,y+d))=g(\mathcal{T}(x,y))+d$.

Now consider a sub-partition of partition iii of random space that has $p$ non-detecting sensors and $n$ detecting sensors. The mean square error for this sub-partition would be,

\begin{eqnarray}
E[\left|\hat{z_{\text{T}}}-z_{\text{T}}\right|^2]=   \binom{p}{1}  \binom{n}{1}   \left(\frac{1}{L}\right)^2 \int_{z_{\text{T}}-3R}^{z_{\text{T}}-R}  \int_{z_{\text{T}}-R}^{z_{\text{T}}+R}  \left(\frac{2R-(y-x)}{L}\right)^{n-1}    \nonumber\\
 \left(\frac{L-2R-(y-x)}{L}\right)^{p-1} \left| g(\mathcal{T}(x,y))-z_{\text{T}} \right|^2  \mathbf{1}_{y-x <2R}   dy dx
\end{eqnarray}
where $\left(\frac{2R-(y-x)}{L}\right)^{n-1} $ is the probability that all other detecting sensors are located inside $[y,x+2R]$ and $\left(\frac{L-2R-(y-x)}{L}\right)^{p-1}$ is the probability of all other non-detecting sensors will be located in $[A_1,x] \cup [y+2R,A_2]$ interval.
Now, with a change of variable $t=y-x$, removing $y$ and considering that $t \in \left[0,~2R \right]$. We have,
\begin{eqnarray}
E[\left|\hat{z_{\text{T}}}-z_{\text{T}}\right|^2]=  \frac{np}{L^2}  \int_{z_{\text{T}}-3R}^{z_{\text{T}}-R}  \int_{0}^{2R}
\left(\frac{2R-t}{L}\right)^{n-1}  \left(\frac{L-2R-t}{L}\right)^{p-1}   \nonumber\\
\left| g(\mathcal{T}(x,x+t))-z_{\text{T}} \right|^2  \mathbf{1}_{z_{\text{T}} \in \left[ x+R, x+t+R \right] }  dt dx
\end{eqnarray}
where $\mathbf{1}_{z_{\text{T}} \in \left[  x+R, x+t+R \right]}$ and $t \in [0,2R]$ guarantees that $y$ will be located in $[z_{\text{T}}-R,z_{\text{T}}+R]$.
The assumption regarding invariance of $g$ to Euclidean motion result in $g(\mathcal{T}(0+x,t+x))=g(\mathcal{T}(0,t))+x$. In addition, $ \mathbf{1}_{z_{\text{T}} \in \left[ x+R, x+t+R \right] }=\mathbf{1}_{z_{\text{T}}-x \in \left[R, t+R\right]}$. Thus, we have,

\begin{eqnarray}
E[\left|\hat{z_{\text{T}}}-z_{\text{T}}\right|^2]= \frac{np}{L^2}  \int_{z_{\text{T}}-3R}^{z_{\text{T}}-R}  \int_{0}^{2R}
\left(\frac{2R-t}{L}\right)^{n-1}  \left(\frac{L-2R-t}{L}\right)^{p-1}    \nonumber\\
 \left| g(\mathcal{T}(0,t))+x-z_{\text{T}} \right|^2 \mathbf{1}_{z_{\text{T}} -x \in \left[R, t+R \right]}  dt dx
\end{eqnarray}

With another change of variable $t'=z_{\text{T}} -x$ and eliminating $x$ and $z_{\text{T}}$ we have,
\begin{eqnarray}
E[\left|\hat{z_{\text{T}}}-z_{\text{T}}\right|^2]=  \frac{np}{L^2}  \int_{R}^{3R}  \int_{0}^{2R}
\left(\frac{2R-t}{L}\right)^{n-1}  \left(\frac{L-2R-t}{L}\right)^{p-1}    \nonumber\\
\left| g(\mathcal{T}(0,t))-t' \right|^2 \mathbf{1}_{t' \in \left[R,t+R\right] }  dt dt'
\end{eqnarray}
Moreover, because the interval of the two integrals are independent, we can change the order of the integrals and obtain,
\begin{eqnarray}
E[\left|\hat{z_{\text{T}}}-z_{\text{T}}\right|^2]=   \frac{np}{L^2}  \int_{0}^{2R}  \int_{R}^{3R}
\left(\frac{2R-t}{L}\right)^{n-1}  \left(\frac{L-2R-t}{L}\right)^{p-1}    \nonumber\\
\left| g(\mathcal{T}(0,t))-t' \right|^2 \mathbf{1}_{t' \in \left[R,t+R\right] }  dt' dt \nonumber\\
= \frac{np}{L^2} \int_{0}^{2R}  \left(\frac{2R-t}{L}\right)^{n-1}  \left(\frac{L-2R-t}{L}\right)^{p-1} \int_{R}^{3R}   \nonumber\\
 \left| g(\mathcal{T}(0,t))-t' \right|^2 \mathbf{1}_{t' \in \left[ R,t+R\right] }  dt' dt
\end{eqnarray}
Furthermore, because $t \in [0,2R]$ we know that $\left[R,~ t+R \right] \subset \left[R,3R \right]$, thus,
\begin{eqnarray}
E[\left|\hat{z_{\text{T}}}-z_{\text{T}}\right|^2]= \frac{np}{L^2} \int_{0}^{2R}  \left(\frac{2R-t}{L}\right)^{n-1}  \left(\frac{L-2R-t}{L}\right)^{p-1} \int_{R}^{t+R}  \nonumber\\
\left| g(\mathcal{T}(0,t))-t' \right|^2
 dt' dt
\end{eqnarray}
Substituting $\mathcal{T}(0,t)$ by its value $[R,R+t]$ we have,
\begin{eqnarray}
\label{LastMSE_S5}
E[\left|\hat{z_{\text{T}}}-z_{\text{T}}\right|^2]= \frac{np}{L^2} \int_{0}^{2R}  \left(\frac{2R-t}{L}\right)^{n-1}  \left(\frac{L-2R-t}{L}\right)^{p-1} \nonumber\\
\left[\int_{R}^{t+R}  \left| g([R,~R+t])-t' \right|^2  dt' \right]  dt
\end{eqnarray}
On the other hand, $g([R,~R+t])$ does not depend on $t'$ and we know that if it is selected as the center of gravity of the inner integration interval i.e.
\begin{eqnarray}
g([R,~t+R])=CG([R,~t+R])=\frac{t+R+R}{2},
\end{eqnarray}
the inner integral will always be minimized \citep{longley2005geographic,Tan2006DataMining,kogan2006grouping}. Hence because the outer integral integrand is always positive, the entire expectation will be minimized.

Furthermore, similar to derivation presented in (\ref{LastMSE_S5}), we can show that
\begin{eqnarray}
\label{LastBiase}
E[\hat{z_{\text{T}}}-z_{\text{T}}]=  \frac{np}{L^2} \int_{0}^{2R}  \left(\frac{2R-t}{L}\right)^{n-1}  \left(\frac{L-2R-t}{L}\right)^{p-1} \nonumber\\
\left[\int_{R}^{t+R}  \left( g([R,~R+t])-t' \right)  dt' \right]  dt
\end{eqnarray}
Therefore, it can also be verified that $g([R,~R+t])=CG([R,~R+t])=\frac{t+2R}{2}$ makes (\ref{LastBiase}) zero i.e. the estimator is unbiased. Considering that the followings are equivalent
\begin{eqnarray}
g([x+R,~x+R+t])=g([R,~R+t])+x=CG([R,~R+t])+x \nonumber\\
=CG([x+R,~x+R+t])
\end{eqnarray}
Then, $g(\mathcal{T}(x,y))=CG(\mathcal{T}(x,y))$ will be the minimum variance unbiased estimator among all estimators that are invariant to Euclidean motion.


\subsection{Unknown Detection Radius}
\label{sec:NonCensoredRUnknown}
In this scenario, we will reconsider the results when $R$ is itself a target of estimation. Similar to  Section \ref{sec:CensoredRUnknown} the most informative sensor locations are ${z_\text{D}}_{L}$, ${z_\text{D}}_{U}$, ${z_\text{ND}}_{L}$ and ${z_\text{ND}}_{U}$. However since $R$ is not known, definition of \textit{possible target region} is not as straight forward as before. Infact in this case, $R$ can admit a range of values if it meets following conditions.
\begin{eqnarray}
R>\frac{{z_\text{D}}_{U}+{z_\text{D}}_{L}}{2}\\
R<\frac{{z_\text{ND}}_{L}+{z_\text{ND}}_{U}}{2}
\end{eqnarray}
Meeting the first condition means that interval $\left[{z_\text{D}}_{U}-R,{z_\text{D}}_{L}+R\right]$ in (\ref{PTRDefDPlusND}) is not empty and the second condition guarantees that the balls around non-detecting sensors ${z_\text{ND}}_{L}$ and ${z_\text{ND}}_{U}$ will not exclude all the interval from being \textit{possible target region}.

\begin{figure}[!t]
\begin{centering}
\includegraphics[width=5 in]{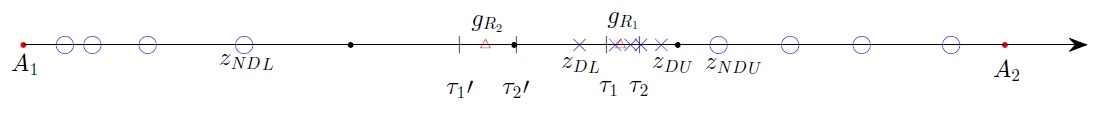}
\caption{The value of $g_R$ for $R_1=0.7$ and $R_2=2.5$ when $({z_\text{D}}_{L},{z_\text{D}}_{U},{z_\text{ND}}_{L},{z_\text{ND}}_{U})=(2.8,3.8,-1.3,4.5)$. }
\label{NonexistenceNoncensoredRUnk}
\end{centering}
\end{figure}

Now, consider a valid observation where all ${z_\text{D}}_{L}$, ${z_\text{D}}_{U}$, ${z_\text{ND}}_{L}$ and ${z_\text{ND}}_{U}$ exist. Depends on what is the actual value of parameter $R$, a different conditions in (\ref{conditions}) is met and $\mathcal{T}$ and $g$ functions would be different. Let us define $g_R\left(\mathbf{Z}_\text{D},\mathbf{Z}_\text{ND}\right)$ as follows:

\begin{eqnarray}
\label{conditionalRrange}
g_R\left(\mathbf{Z}_\text{D},\mathbf{Z}_\text{ND}\right)=\left\{ \begin{array}{lcl}
\frac{ {z_\text{D}}_{U}+ {z_\text{D}}_{L} }{2} & \mbox{ if } &  \left(R \leq \frac{{z_\text{D}}_{U}-{z_\text{ND}}_{L}}{2} \right) \& \left( R \leq \frac{{z_\text{ND}}_{U}-{z_\text{D}}_{L}}{2}  \right)   \\
\frac{ {z_\text{ND}}_{L}+ {z_\text{ND}}_{U}   }{2} & \mbox{ if } & \left(R > \frac{{z_\text{D}}_{U}-{z_\text{ND}}_{L}}{2} \right) \& \left( R > \frac{{z_\text{ND}}_{U}-{z_\text{D}}_{L}}{2}  \right)  \\
\frac{ {z_\text{ND}}_{L} +{z_\text{D}}_{L} }{2}+R & \mbox{ if } & \left(R > \frac{{z_\text{D}}_{U}-{z_\text{ND}}_{L}}{2} \right) \& \left( R \leq \frac{{z_\text{ND}}_{U}-{z_\text{D}}_{L}}{2}  \right)   \\
\frac{ {z_\text{D}}_{U} +{z_\text{ND}}_{U} }{2}+R  & \mbox{ if } & \left(R \leq \frac{{z_\text{D}}_{U}-{z_\text{ND}}_{L}}{2} \right) \& \left( R > \frac{{z_\text{ND}}_{U}-{z_\text{D}}_{L}}{2}  \right)   \\
\end{array} \right.
\end{eqnarray}
Let us consider two different values for $R$, $R_1$ and $R_2$ such that each of them enable a different condition in (\ref{conditionalRrange}). According to Section \ref{sec:NonCensoredRknown} $g_{R_1}\left(\mathbf{Z}_\text{D},\mathbf{Z}_\text{ND}\right)$ and $g_{R_2}\left(\mathbf{Z}_\text{D},\mathbf{Z}_\text{ND}\right)$ would be the MVU estimator when $R=R_1$ and $R=R_2$ respectively but they might have different values as illustrated in Figure \ref{NonexistenceNoncensoredRUnk}. Thus, an MVU estimator does not exist for $R$ unknown case even if we restrict the function space to those Invariant to Euclidean motion.

\section{Conclusion}
\label{sec:Conclusion}
We considered the problem of localization with binary ideal RSS measurements in this paper. Finding a Minimum Variance Unbiased estimator for this scenario is important because it establishes a lower bound for all other estimators in the presence of uncertainty. We showed that even though in the theoretical statistics literature a version of the problem has been categorized as a problem that does not admit any MVU estimator, the result might be different when other characteristics of the estimators are considered. We add the assumption of non-invariance of the estimator under Euclidean motion and show that when the detection radius is known the MVU estimator does exist for both censored and non-censored scenarios under this constraint, whereas it still does not exist for the non-censored scenario when the detection radius is unknown. We believe that this constraint is critical for many other applied estimators, thus care is necessary when applying pure theoretical statistical results to estimation of physical phenomenon. Hence, it may justify the necessity of re-evaluation of some of these results this assumption (or perhaps other ones too) regarding the functional space.

\appendix

\section{proof for condition i}
This condition means that $\mathcal{T}(\mathbf{Z}_\text{D},\mathbf{Z}_\text{ND})=\left[{z_\text{D}}_{U}-R,{z_\text{D}}_{L}+R\right]$. In this sub-section, let us denote ${z_\text{D}}_{L}$ and ${z_\text{D}}_{U}$ by $x$ and $y$ respectively. The \textit{possible target region} would be $\mathcal{T}(\mathbf{Z}_\text{D},\mathbf{Z}_\text{ND})=\mathcal{T}(x,y)=\left[y-R,x+R\right]$. Moreover, under this condition we have that

\begin{eqnarray}
{z_\text{ND}}_{L} \le{z_\text{D}}_{U}-2R \nonumber\\
{z_\text{ND}}_{U} \ge{z_\text{D}}_{L}+2R
\end{eqnarray}
Therefore, all non-detecting sensors should be located in interval $[A_1,{z_\text{D}}_{U}-2R]$ or $[{z_\text{D}}_{L}+2R,A_2]$. An observation $(x,y)$ is valid if and only if $z_{\text{T}} \in [y-R,x+R] $.

As explained in Section \ref{sec:CensoredRKnown}, we can restrict our search for an MVU estimator to functions of the minimal sufficient statistics i.e. $g(\mathcal{T}(\left(x,y\right)))=g(\left[y-R,x+R\right])$. The invariance to Euclidean motion assumption suggests that $g(\mathcal{T}(x+d,y+d))=g(\mathcal{T}(x,y))+d$.

Now consider a sub-partition of random space that has $p$ non-detecting sensors and $n$ detecting sensors. Let us assume that $n>1$. We will examine the result when $n=1$ at the end of this sub-section separately. We can write the Mean Squared Error as following:

\begin{eqnarray}
E[\left|\hat{z_{\text{T}}}-z_{\text{T}}\right|^2]=  \frac{2}{(L)^2} \binom{n}{2}  \int_{z_{\text{T}}-R}^{z_{\text{T}}+R}  \int_{z_{\text{T}}-R}^{z_{\text{T}}+R} \left(\frac{y-x}{L}\right)^{n-2}  \left(\frac{L-4R+y-x}{L}\right)^p  \nonumber\\
\left| g(\mathcal{T}(x,y))-z_{\text{T}} \right|^2  \mathbf{1}_{x \leq y}   dy dx
\end{eqnarray}
where the binomial coefficient compensates for the fact that any two of the $n$ detecting sensors can be selected as the maximum and minimum values of the locations of the detecting sensors and $\left(\frac{y-x}{L}\right)^{n-2}$ is the probability that the other detecting sensors are located inside $[x,y]$; and $\left(\frac{L-4R+y-x}{L}\right)^p$ is the probability of all non-detecting sensors will be located in $[A_1,y-2R] \cup [x+2R,A_2]$ interval.
Now, with a change of variable $t=y-x$, removing $y$ and considering that $t \in \left[0,2R \right]$. We have,

\begin{eqnarray}
E[\left|\hat{z_{\text{T}}}-z_{\text{T}}\right|^2]=  \frac{n(n-1)}{L^2}  \int_{z_{\text{T}}-R}^{z_{\text{T}}+R}  \int_{0}^{2R} \left(\frac{t}{L}\right)^{n-2}  \left(\frac{L-4R+t}{L}\right)^p  \nonumber\\
\left| g(\mathcal{T}(x,x+t))-z_{\text{T}} \right|^2  \mathbf{1}_{z_{\text{T}} \in \left[ t+x-R,R+x \right] }  dt dx
\end{eqnarray}
where $\mathbf{1}_{z_{\text{T}} \in \left[ t+x-R,R+x \right]}$ guarantees that $x$ and $x+t$ will be located in the detecting area.
The assumption regarding invariance of $g$ to Euclidean motion result in $g(\mathcal{T}(0+x,t+x))=g(\mathcal{T}(0,t))+x$. In addition, $\mathbf{1}_{z_{\text{T}}\in \left[ t+x-R,R+x \right]}=\mathbf{1}_{z_{\text{T}}-x \in \left[ t-R,R\right]}$. Thus, we have,

\begin{eqnarray}
E[\left|\hat{z_{\text{T}}}-z_{\text{T}}\right|^2]=  \frac{n(n-1)}{L^2}  \int_{z_{\text{T}}-R}^{z_{\text{T}}+R}  \int_{0}^{2R} \left(\frac{t}{L}\right)^{n-2} \left(\frac{L-4R+t}{L}\right)^p   \nonumber\\
\left| g(\mathcal{T}(0,t))+x-z_{\text{T}} \right|^2 \mathbf{1}_{z_{\text{T}} -x \in \left[ t-R,R\right]}  dt dx
\end{eqnarray}
With another change of variable $t'=z_{\text{T}} -x$ and eliminating $x$ and $z_{\text{T}}$ we have,
\begin{eqnarray}
E[\left|\hat{z_{\text{T}}}-z_{\text{T}}\right|^2]= \frac{n(n-1)}{L^2}  \int_{-R}^{+R}  \int_{0}^{2R} \left(\frac{t}{L}\right)^{n-2} \left(\frac{L-4R+t}{L}\right)^p  \left| g(\mathcal{T}(0,t))-t' \right|^2  \nonumber\\
 \mathbf{1}_{t' \in \left[ t-R,R\right] }  dt dt'
\end{eqnarray}
Moreover, because the interval of the two integrals are independent, we can change the order of the integrals and obtain,
\begin{eqnarray}
E[\left|\hat{z_{\text{T}}}-z_{\text{T}}\right|^2]=
\frac{n(n-1)}{L^2}   \int_{0}^{2R}   \left(\frac{t}{L}\right)^{n-2}  \left(\frac{L-4R+t}{L}\right)^p \int_{-R}^{+R} \nonumber\\
\left| g(\mathcal{T}(0,t))-t' \right|^2   \mathbf{1}_{t'\in \left[ t-R,R\right] }  dt' dt
\end{eqnarray}
Furthermore, because $0 \leq t \leq 2R$ we know that $\left[ t-R,~ R \right] \subset \left[  -R, ~ R \right]$, thus,
\begin{eqnarray}
E[\left|\hat{z_{\text{T}}}-z_{\text{T}}\right|^2]=  \frac{n(n-1)}{L^2}   \int_{0}^{2R}   \left(\frac{t}{L}\right)^{n-2}  \left(\frac{L-4R+t}{L}\right)^p \nonumber\\
\left[  \int_{t-R}^{R}  \left| g(\mathcal{T}(0,t))-t' \right|^2  dt' \right]  dt
\end{eqnarray}
Substituting $\mathcal{T}(0,t)$ by its value $[t-R,R]$ we have,
\begin{eqnarray}
\label{LastMSE}
E[\left|\hat{z_{\text{T}}}-z_{\text{T}}\right|^2]=  \frac{n(n-1)}{L^2}   \int_{0}^{2R}  \left(\frac{t}{L}\right)^{n-2}  \left(\frac{L-4R+t}{L}\right)^p \nonumber\\
\left[  \int_{t-R}^{R}  \left| g([t-R,~R])-t' \right|^2  dt' \right]  dt
\end{eqnarray}
On the other hand, $g([t-R,~R])$ does not depend on $t'$ and we know that if it is selected as the center of gravity of the inner integration interval i.e.
\begin{eqnarray}
g([t-R,~R])=CG([t-R,R])=\frac{t-R+R}{2},
\end{eqnarray}
the inner integral will always be minimized \citep{longley2005geographic,Tan2006DataMining,kogan2006grouping}. Hence because the outer integral integrand is always positive, the entire expectation will be minimized.

Furthermore, similar to derivation presented in (\ref{LastMSE}), we can show that
\begin{eqnarray}
\label{LastBiaseC1}
E[\hat{z_{\text{T}}}-z_{\text{T}}]=  \frac{n(n-1)}{L^2}   \int_{0}^{2R}   \left(\frac{t}{L}\right)^{n-2}  \left(\frac{L-4R+t}{L}\right)^p \nonumber\\
\left[  \int_{t-R}^{R}  \left( g([t-R,~R])-t' \right)  dt' \right]  dt
\end{eqnarray}
Therefore, it also can be verified that $g([t-R,~R])=CG([t-R,R])=\frac{t-R+R}{2}$ makes (\ref{LastBiaseC1}) zero i.e. the estimator is unbiased. Considering that the followings are equivalent
\begin{eqnarray}
g([t+x-R,~x+R])=g([t-R,~R])+x= CG([t-R,R])+x\nonumber\\
=CG([x+t-R,x+R])
\end{eqnarray}
Then, $g(\mathcal{T}(x,y))=CG(\mathcal{T}(x,y))$ will be the minimum variance unbiased estimator among all estimators that are invariant to Euclidean motion.
If $n=1$, the Mean Squared Error would become
\begin{eqnarray}
E[\left|\hat{z_{\text{T}}}-z_{\text{T}}\right|^2]=\frac{1}{L} \int_{z_{\text{T}}-R}^{z_{\text{T}}+R} \left(\frac{L-4R}{L}\right)^p \left| g(\mathcal{T}(x,x))-z_{\text{T}} \right|^2 dx
\end{eqnarray}
with a change of variable $t'=z_{\text{T}}-x$ and similar effort as previously presented, we will get that,
\begin{eqnarray}
E[\left|\hat{z_{\text{T}}}-z_{\text{T}}\right|^2]=\frac{1}{L} \int_{-R}^{+R} \left(\frac{L-4R}{L}\right)^p \left| g(\mathcal{T}(0,0))-t' \right|^2 dt'
\end{eqnarray}
Therefore, $g(\mathcal{T}(0,0))=0$ make the MSE minimized and therefore $g(\mathcal{T}(x,x))=x$ will be the MVU estimator.


\section{proof for condition ii}
This condition means that $\mathcal{T}(\mathbf{Z}_\text{D},\mathbf{Z}_\text{ND})=\left[{z_\text{ND}}_{L}+R,{z_\text{ND}}_{U}-R\right]$. In this sub-section, let us denote ${z_\text{ND}}_{L}$ and ${z_\text{ND}}_{U}$ by $x$ and $y$ respectively. The \textit{possible target region} would be $\mathcal{T}(\mathbf{Z}_\text{D},\mathbf{Z}_\text{ND})=\mathcal{T}(\left(x,y\right))=\left[x+R,y-R\right]$. Moreover,
\begin{eqnarray}
{z_\text{D}}_{U} < {z_\text{ND}}_{L}+2R  \nonumber\\
{z_\text{D}}_{L} > {z_\text{ND}}_{U}-2R
\end{eqnarray}
Therefore, all detecting sensors should be located in interval $(y-2R,x+2R)$ and all non-detecting sensors should be located in interval $[A_1,x]$ or $[y,A_2]$. In this case, an observation $(x,y)$ implies that $z_{\text{T}} \in [x+R,y-R] $. Note that in order for $[y-2R,x+2R]$ to be a valid interval it is implicit that $0<y-x<4R$.

As explained in Section \ref{sec:CensoredRKnown}, we can restrict our search for an MVU estimator to functions of the minimal sufficient statistics i.e. $g(\mathcal{T}(\left(x,y\right)))=g(\left[x+R,y-R\right])$. The invariance to Euclidean motion assumption suggests that $g(\mathcal{T}(x+d,y+d))=g(\mathcal{T}(x,y))+d$.

Now consider a sub-partition of random space that has $p$ non-detecting sensors and $n$ detecting sensors. We can write the Mean Square Error as following:

\begin{eqnarray}
E[\left|\hat{z_{\text{T}}}-z_{\text{T}}\right|^2]=  \frac{2}{L^2} \binom{p}{2}   \int_{z_{\text{T}}-3R}^{z_{\text{T}}-R}  \int_{z_{\text{T}}+R}^{z_{\text{T}}+3R} \left(\frac{4R+x-y}{L}\right)^{n}  \left(\frac{L-(y-x)}{L}\right)^{p-2}  \nonumber\\
\left| g(\mathcal{T}(x,y))-z_{\text{T}} \right|^2  \mathbf{1}_{y-x \leq 4R}   dy dx.
\end{eqnarray}
where the binomial coefficient compensates for the fact that any two of the $p$ non-detecting sensors can be ${z_\text{ND}}_{L}$ and ${z_\text{ND}}_{U}$. Moreover, $\left(\frac{4R-(y-x)}{L}\right)^{n} $ is the probability that all detecting sensors are located inside $[y-2R,x+2R]$ and $\left(\frac{L-(y-x)}{L}\right)^{p-2}$ is the probability of all other non-detecting sensors will be located in $[A_1,x] \cup [y,A_2]$ interval.
Now, with a change of variable $t=y-x$, removing $y$ and considering that $t \in \left[2R,4R \right]$, we have,
\begin{eqnarray}
E[\left|\hat{z_{\text{T}}}-z_{\text{T}}\right|^2]=  \frac{p(p-1)}{L^2}  \int_{z_{\text{T}}-3R}^{z_{\text{T}}-R}  \int_{2R}^{4R} \left(\frac{4R-t}{L}\right)^{n}  \left(\frac{L-t}{L}\right)^{p-2}   \nonumber\\
\left| g(\mathcal{T}(x,x+t))-z_{\text{T}} \right|^2 \mathbf{1}_{z_{\text{T}} \in \left[ x+R, x+t-R \right] }  dt dx.
\end{eqnarray}
where $\mathbf{1}_{z_{\text{T}} \in \left[ x+R,x+t-R \right]}$ guarantees that $x$ and $x+t$ will be located in the non-detecting area.
The assumption regarding invariance of $g$ to Euclidean motion result in $g(\mathcal{T}(0+x,t+x))=g(\mathcal{T}(0,t))+x$. In addition, $ \mathbf{1}_{z_{\text{T}} \in \left[ x+R, x+t-R \right] }=\mathbf{1}_{z_{\text{T}}-x \in \left[ R,t-R\right]}$. Thus, we have,

\begin{eqnarray}
E[\left|\hat{z_{\text{T}}}-z_{\text{T}}\right|^2]=  \frac{p(p-1)}{L^2}  \int_{z_{\text{T}}-3R}^{z_{\text{T}}-R}  \int_{2R}^{4R} \left(\frac{4R-t}{L}\right)^{n}  \left(\frac{L-t}{L}\right)^{p-2}    \nonumber\\
\left| g(\mathcal{T}(0,t))+x-z_{\text{T}} \right|^2  \mathbf{1}_{z_{\text{T}} -x \in \left[ R,t-R\right]}  dt dx.
\end{eqnarray}

With another change of variable $t'=z_{\text{T}} -x$ and eliminating $x$ and $z_{\text{T}}$ we have,
\begin{eqnarray}
E[\left|\hat{z_{\text{T}}}-z_{\text{T}}\right|^2]=  \frac{p(p-1)}{L^2}  \int_{R}^{3R}  \int_{2R}^{4R} \left(\frac{4R-t}{L}\right)^{n}  \left(\frac{L-t}{L}\right)^{p-2}   \nonumber\\
\left| g(\mathcal{T}(0,t))-t' \right|^2 \mathbf{1}_{t' \in \left[ R,t-R\right] }  dt dt'
\end{eqnarray}
Moreover, because the interval of the two integrals are independent, we can change the order of the integrals and obtain,
\begin{eqnarray}
E[\left|\hat{z_{\text{T}}}-z_{\text{T}}\right|^2]= \frac{p(p-1)}{L^2}   \int_{2R}^{4R}     \left(\frac{4R-t}{L}\right)^{n}  \left(\frac{L-t}{L}\right)^{p-2}  \int_{R}^{3R}   \nonumber\\
\left| g(\mathcal{T}(0,t))-t' \right|^2 \mathbf{1}_{t'\in \left[ R,t-R\right] }  dt' dt
\end{eqnarray}

Furthermore, because $t \in [2R,4R]$ we know that $\left[ R,~ t-R \right] \subset \left[  R, ~ 3R \right]$, thus,
\begin{eqnarray}
E[\left|\hat{z_{\text{T}}}-z_{\text{T}}\right|^2]=  \frac{p(p-1)}{L^2}   \int_{2R}^{4R}    \left(\frac{4R-t}{L}\right)^{n}  \left(\frac{L-t}{L}\right)^{p-2} \nonumber\\
\left[  \int_{R}^{t-R}    \left| g(\mathcal{T}(0,t))-t' \right|^2  dt' \right]  dt
\end{eqnarray}
Substituting $\mathcal{T}(0,t)$ by its value $[R,t-R]$ we have,
\begin{eqnarray}
\label{LastMSE_S2}
E[\left|\hat{z_{\text{T}}}-z_{\text{T}}\right|^2]=  \frac{p(p-1)}{L^2}   \int_{2R}^{4R}   \left(\frac{4R-t}{L}\right)^{n}  \left(\frac{L-t}{L}\right)^{p-2} \nonumber\\
\left[  \int_{R}^{t-R}  \left| g([R,~t-R])-t' \right|^2  dt' \right]  dt
\end{eqnarray}
On the other hand, $g([R,~t-R])$ does not depend on $t'$ and we know that if it is selected as the center of gravity of the inner integration interval i.e.
\begin{eqnarray}
g([R,~t-R])=CG([R,~t-R])=\frac{t-R+R}{2},
\end{eqnarray}
the inner integral will always be minimized \citep{longley2005geographic,Tan2006DataMining,kogan2006grouping}. Hence, because the outer integral integrand is always positive\footnote{Note that $L-t\ge0$ because of the assumptions that the deployment is so large that $[z_{\text{T}}-3R,z_{\text{T}}+3R] \subset [A_1,A_2]$}, the entire expectation will be minimized.

Furthermore, similar to derivation presented in (\ref{LastMSE_S2}), we can show that
\begin{eqnarray}
\label{LastBiaseS2}
E[\hat{z_{\text{T}}}-z_{\text{T}}]=  \frac{p(p-1)}{L^2}   \int_{2R}^{4R}   \left(\frac{4R-t}{L}\right)^{n}  \left(\frac{L-t}{L}\right)^{p-2} \nonumber\\
\left[  \int_{R}^{t-R}  \left( g([R,~t-R])-t' \right)  dt' \right]  dt
\end{eqnarray}
It also can be verified that $g([R,~t-R])=CG([R,~t-R])=\frac{t-R+R}{2}$ makes (\ref{LastBiaseS2}) zero i.e. the estimator is unbiased. Considering that the followings are equivalent
\begin{eqnarray}
g([x+R,~x+t-R])=g([R,~t-R])+x= CG([R,~t-R])+x\nonumber\\
=CG([x+R,~x+t-R])
\end{eqnarray}
Then, $g(\mathcal{T}(x,y))=CG(\mathcal{T}(x,y))$ will be the minimum variance unbiased estimator among all estimators that are invariant to Euclidean motion.

\section{proof for condition iv}

Conditions iv) would be similar to conditions iii) because of symmetry. Therefore, we will not repeat the proof here for the sake of space.

\bibliographystyle{IEEEbib}
\bibliography{refs}

\newpage

\end{document}